\documentclass[prl,twocolumn,superscriptaddress,showpacs]{revtex4}

\usepackage{graphicx}

\usepackage[normalem]{ulem} 
\usepackage{color} 

\begin{document}

\title{Strong coupling of an Er$^{3+}$ doped YAlO$_{3}$ crystal to a superconducting resonator}

\author{A.~Tkal\v{c}ec}
\affiliation{Physikalisches Institut, Karlsruhe Institute of Technology, D-76128 Karlsruhe, Germany}

\author{S.~Probst}
\affiliation{Physikalisches Institut, Karlsruhe Institute of Technology, D-76128 Karlsruhe, Germany}

\author{D.~Rieger}
\affiliation{Physikalisches Institut, Karlsruhe Institute of Technology, D-76128 Karlsruhe, Germany}

\author{H.~Rotzinger}
\affiliation{Physikalisches Institut, Karlsruhe Institute of Technology, D-76128 Karlsruhe, Germany}

\author{S.~W\"{u}nsch}
\affiliation{Institut f\"{u}r Mikro- und Nanoelektronische Systeme, Karlsruhe Institute of Technology, D-76189 Karlsruhe, Germany}

\author{N.~Kukharchyk}
\affiliation{Angewandte Festk\"{o}rperphysik, Ruhr-Universit\"{a}t Bochum, D-44780 Bochum, Germany}

\author{A.~D.~Wieck}
\affiliation{Angewandte Festk\"{o}rperphysik, Ruhr-Universit\"{a}t Bochum, D-44780 Bochum, Germany}

\author{M.~Siegel}
\affiliation{Institut f\"{u}r Mikro- und Nanoelektronische Systeme, Karlsruhe Institute of Technology, D-76189 Karlsruhe, Germany}

\author{A.~V.~Ustinov}
\affiliation{Physikalisches Institut, Karlsruhe Institute of Technology, D-76128 Karlsruhe, Germany}

\author{P.~Bushev}
\affiliation{Experimentalphysik, Universit\"{a}t des Saarlandes, D-66123 Saarbr\"{u}cken, Germany}

\date{\today}

\begin{abstract}
Quantum memories are integral parts of both quantum computers and quantum communication networks. Naturally, such a memory is embedded into a hybrid quantum architecture, which has to meet the requirements of fast gates, long coherence times and long distance communication. Erbium doped crystals are well suited as a microwave quantum memory for superconducting circuits with additional access to the optical telecom C-band around 1.55 $\mu$m. Here, we report on circuit QED experiments with an Er$^{3+}$:YAlO$_{3}$ crystal and demonstrate strong coupling to a superconducting lumped element resonator. The low magnetic anisotropy of the host crystal allows for attaining the strong coupling regime at relatively low magnetic fields, which are compatible with superconducting circuits. In addition, Ce$^{3+}$ impurities were detected in the crystal, which showed strong coupling as well. \end{abstract}



\maketitle


Reliable operation of quantum information and communication protocols requires a quantum memory (QM), i.e. a system, which allows for storage and on-demand retrieval of a quantum bit~\cite{Gisin2007, Simon2010}. This can be realized by a great variety of physical systems such as single trapped ions~\cite{Monroe2004}, atoms~\cite{Weinfurter2007}, single spins~\cite{Maurer2012}, two-level defects~\cite{Neeley2008} and spin-ensembles~\cite{Steger2012}, which differ by their frequency band, coherence time and operating conditions.

Rare-earth (RE) ions doped into a solid represent one of most promising systems suitable for quantum memories, because their inner shell 4f optical electronic transitions possess very long coherence times~\cite{Thiel2011}. The excellent optical properties of RE doped crystals are confirmed and harvested by the world wide research in quantum optics. This includes a light-matter interface at the single photon level~\cite{Riedmatten2008}, an efficient and broadband quantum memory for light~\cite{Sellars2010, Tittel2011}, a quantum memory at the telecom C-band~\cite{Gisin2010}, an atomic frequency comb memory~\cite{Kroel2010}, storage of entanglement in a RE doped crystal~\cite{Clausen2011} and generation of entanglement between two crystals~\cite{Usmani2012}. Yet, in contrast to the single atom approach, a quantum memory based on RE doped solids allows for the implementation of multimode storage protocols~\cite{Nunn2008, Afzelius2009}.

There are seven RE's ions (Ce$^{3+}$, Nd$^{3+}$, Sm$^{3+}$, Gd$^{3+}$, Dy$^{3+}$, Er$^{3+}$, Yb$^{3+}$), which are suited for a microwave quantum memory due to the presence of a large electronic spin associated with an unquenched orbital moment~\cite{AbragamESR}. Most of them have access to the nuclear spin degrees of freedom, which allow for long term storage~\cite{Bertaina2007}. These RE ions can be doped into a variety of host crystals, and therefore, can potentially be integrated with superconducting (SC) quantum circuits~\cite{Imamoglu2009}. The resulting hybrid quantum system can consist of a SC qubit, a transmission line or a resonator magnetically coupled to the spin ensemble~\cite{Marcos2010}. The exclusive feature of some RE ions (Nd$^{3+}$, Er$^{3+}$, Yb$^{3+}$) is the presence of optical transitions inside standard telecommunication bands. A quantum memory based on these RE elements can be very attractive for quantum communication between qubits of different physical nature~\cite{Rabl2010}. Particularly in the case of SC qubits, the RE spin ensemble is expected to act as a coherent and reversible quantum converter between the optical domain and the microwave frequency band~\cite{Bushev2011, Afzelius2013, Nemoto2013}.

The research on microwave quantum memories has started few years ago and is primarily focused on circuit QED experiments with electronic spin ensembles of nitrogen vacancies (NV) in diamond. The strong coupling of spins to SC-resonators~\cite{Bertet2010, Shuster2010, Amsuss2011, DiCarlo2013} as well as the full hybrid quantum architecture has recently been demonstrated~\cite{Bertet2011, Saito2013}. The exploration of hybrid systems based on Er$^{3+}$:Y$_2$SiO$_5$~(Er:YSO) crystals coupled to SC resonators has started successfully~\cite{Bushev2011, Wilson2012}. Such a crystal is known for the longest measured optical coherence time among solid state systems~\cite{Sun2006} and for its large spin tuning rate of up to 200 GHz/T~\cite{Kurkin1980,Guillot2006,Sun2008}. We found that in order to implement an Er:YSO based QM at the microwave C-band (4-8 GHz), a large magnetic field above 200 mT is required~\cite{Probst2013}. The magnetic anisotropy of Er:YSO demands trading of the coupling strength for the high spin tuning rate. However, since our goal is the coherent integration into the state of the art circuit QED architecture, the operation at lower magnetic fields is desirable~\cite{Bothner2011}.

In this article, we report on circuit QED experiments with an Er$^{3+}$:YAlO$_{3}$ (Er:YAP) crystal and demonstrate strong coupling at a relatively small magnetic field of 33~mT. Ce$^{3+}$ impurities, which were detected in the crystal, showed strong coupling, too.

\begin{figure}[htb]
\includegraphics[width=0.9\columnwidth]{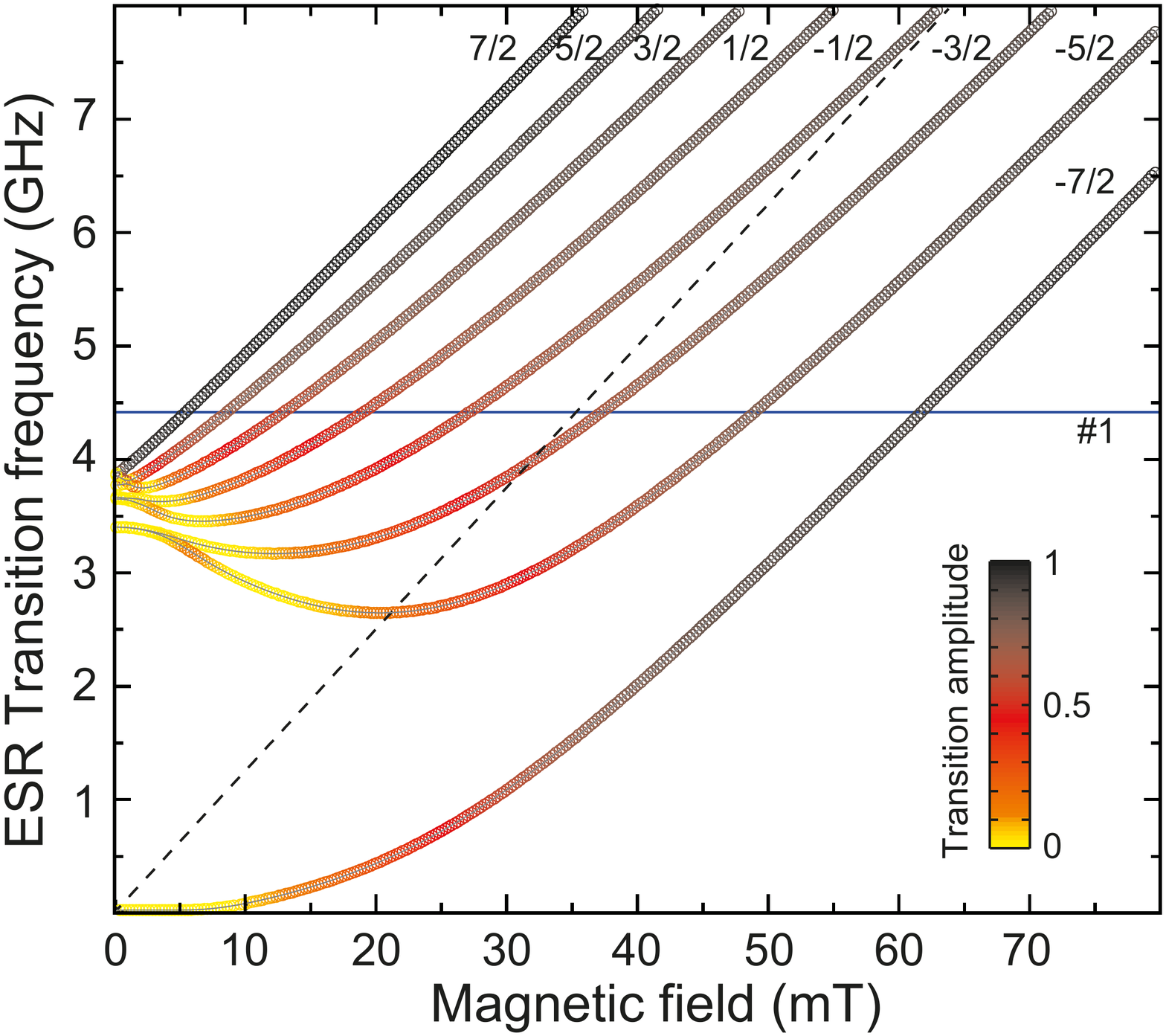}
\caption{(Color online)~Simulated ESR spectrum of the Er:YAP crystal. The color bar represents the amplitude of the HF transitions of $^{167}$Er (23\% abundance). The dashed line shows the electronic spin transition of the even isotopes. The horizontal blue line indicated by \#1 denotes the resonance frequency of the superconducting LE resonator used in our experiment. The HF spin transitions are labeled accordingly by their nuclear spin projection number $m_I$.}
\label{fig_hf_simulation}
\end{figure}

Rare earth doped YAP crystals belong to the orthorombic $D_{2h}$ space group and reveal a relatively low magnetic anisotropy compared to YSO~\cite{Asatryan1997}, which has already been shown to be suitable for the integration with SC quantum circuits \cite{Bushev2011, Wilson2012}. The unit cell of a YAP crystal consists of four distorted perovskite cells creating four magnetic classes for the Y$^{3+}$ ions. However, due the mirror symmetry of the yttrium in the (001) plane and the inversion symmetry through the aluminum sites, only two magnetically inequivalent positions remain. Thus, one expects in the ESR spectrum to see maximum two electronic spin transitions. For the RE isotopes with a nuclear spin ($^{167}$Er with $I=7/2$), magnetic transitions between the hyperfine (HF) levels can be observed.

In order to understand the essential features of the electron spin resonance (ESR) spectrum of Er:YAP, we numerically diagonalize the spin Hamiltonian $H = \mu_B \textbf{B} \cdot \textbf{g} \cdot \textbf{S} + \textbf{I}\cdot \textbf{A}\cdot \textbf{S}$ using the EASYSPIN~\cite{EasySpin} software according to our experimental setting. The first term of the Er:YAP Hamiltonian represents the electronic Zeeman splitting, whereas the second one describes the hyperfine interactions. The values for the \textbf{\textrm{g}}-tensor and the HF tensor \textbf{A} were taken from~\cite{Asatryan1997}. The resulting frequency spectrum is presented in Fig.~\ref{fig_hf_simulation}. The straight dashed line crossing the spectrum corresponds to the electronic spin transition between the states $m_s=\pm1/2$. The other lines correspond to the hyperfine transitions of the $^{167}$Er isotope with preserved nuclear magnetic number $m_I$. The color of each HF line indicates the calculated amplitude of the transition. The $^{167}$Er:YAP system has a zero-field splitting around 4 GHz, which could allow for the operation of a flux qubit coupled directly to the spin ensemble at zero magnetic field~\cite{Marcos2010, Zhu2011}. Another interesting feature is the presence of so called zero first-order Zeeman (ZEFOZ) transitions. These can greatly improve the spin dephasing time due to their first order insensitivity to magnetic field fluctuations~\cite{Longdell2012}.

\begin{figure}[htb]
\includegraphics[width=0.95\columnwidth]{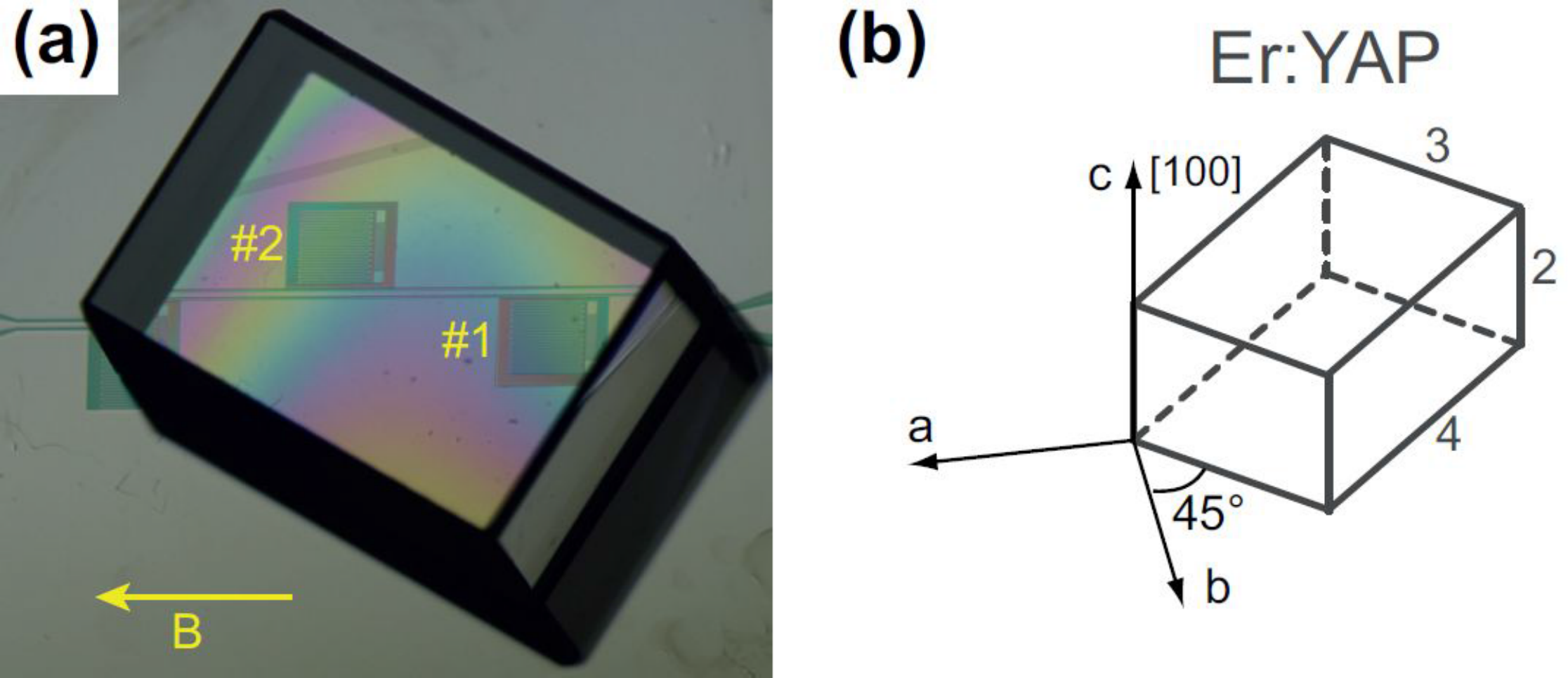}
\caption{(Color online)~\textbf{(a)}~Picture of the hybrid system. The Er:YAP crystal of dimension 2 $\times$ 3 $\times$ 4 mm$^3$ is placed on the niobium chip with 3 LE resonators. The magnetic field is applied parallel to the chip's surface coinciding with the (001) plane of the crystal. The crystal occupies the mode volumes of resonators \#1 and \#2.~\textbf{(b)}~The orientation of the crystal's axis is displayed by using $P_{nma}$ notation. The dimensions of the crystal are given in mm.}
\label{fig_chip_and_crystal}
\end{figure}

In this experiment, we use a single YAlO$_{3}$ crystal doped with 0.005\% of Er$^{3+}$, supplied by Scientific Materials Inc. The concentration of erbium spins in the YAP crystal is similar to the dopant concentration of Er:YSO used in our previous experiments and estimated to be $n_s\simeq10^{18}$~cm$^{-3}$~\cite{Probst2013}. The crystal has dimensions of 2~$\times$~3 $\times$~4 mm$^3$ and is placed onto a SC niobium chip with three lumped element~(LE) resonators~\cite{Wuensch2011}. Figure.~\ref{fig_chip_and_crystal}~(a) shows a micrograph of the chip, where the three LE resonators are coupled to a common 50 Ohm transmission line. The design of an individual resonator is similar to the ones used in our previous study~\cite{Probst2013}. The resonance frequencies of the bare LE chip occupy the frequency band between 5 and 5.5 GHz. After putting the crystal on top of the resonators, their frequencies shift down to approximately 4.4 GHz due to the permittivity of YAlO$_{3}$. In this setting, the LE resonators reveal intrinsic and loaded quality factors of $Q_i\simeq$~2900 and $Q_l\simeq$~540, respectively.

The orientation of the crystal on the chip is shown in Fig.~\ref{fig_chip_and_crystal}~(b) denoted in $P_{nma}$ notation. The magnetic field is applied along the surface of the SC chip, which also coincides with the $ab$ plane of the crystal. The $c$-axis matches the [001] direction. The angle between the $b$-axis and the magnetic field was chosen to be 10$^{\circ}$, which optimizes the DC \textrm{g}-factor. The observation of Newton rings confirms that the crystal is in close proximity to the chip and occupies the mode volumes of resonators \#1 and \#2.

\begin{figure}[htb]
\includegraphics[width=0.95\columnwidth]{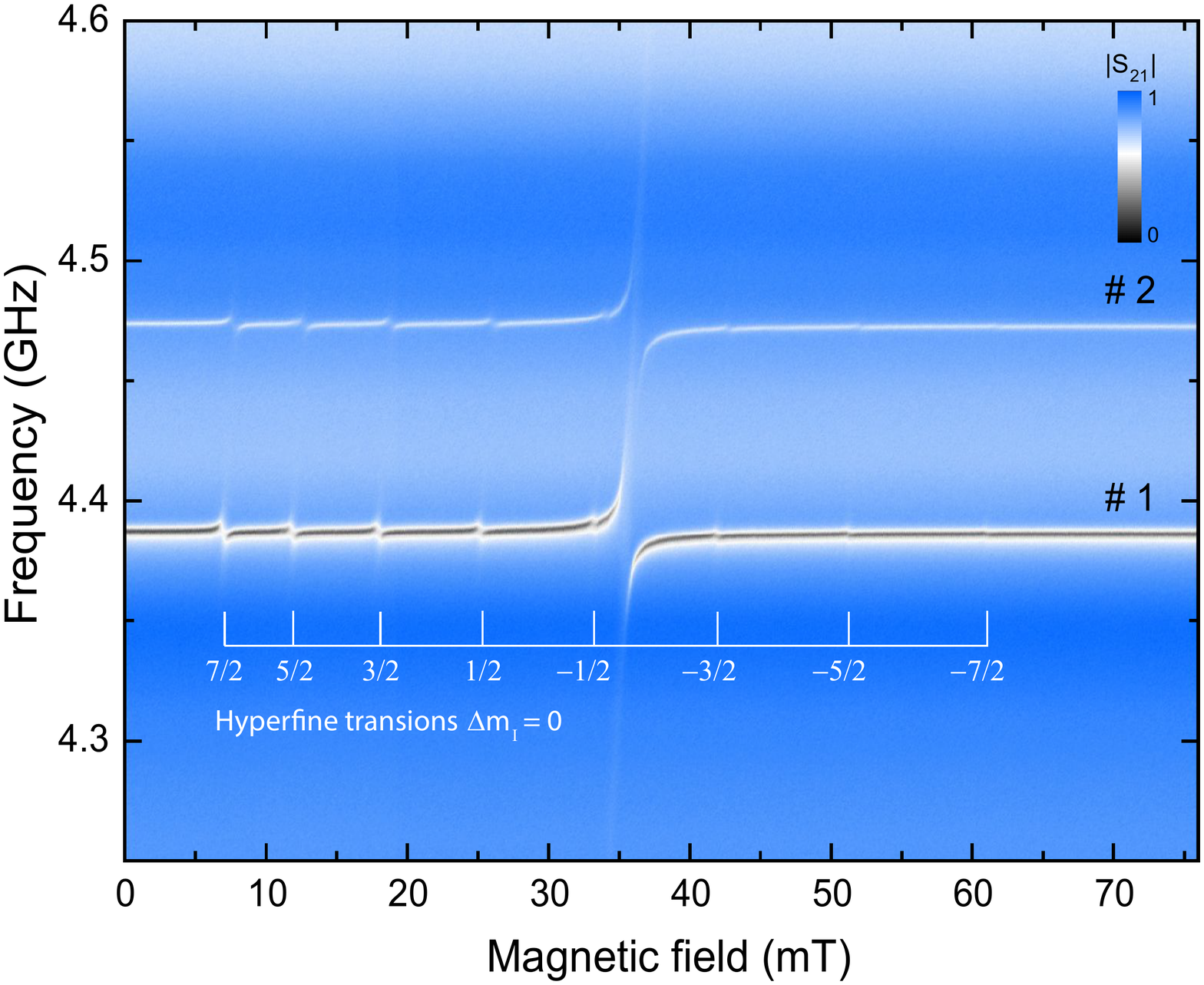}
\caption{(Color online)~ESR sprectrum of the Er:YAP crystal coupled to the LE chip. Two resonators fall into the frequency region of our interest, denoted by \#1 and \#2. 8 HF transitions of $^{167}$Er between the states with nuclear spin number $m_I=7/2, 5/2,$ ... $-7/2$ are clearly resolved and identified.}
\label{fig_er_largespec_and_hf}
\end{figure}

The experiment is placed inside a BlueFors LD-250 dilution fridge with a base temperature of 25 mK. The on-chip ESR spectroscopy is carried out in the common way by measuring the $S_{21}$ microwave transmission of the SC chip and varying the magnetic field, see Refs.~\cite{Bertet2010, Bushev2011, Probst2013, Tobar2013}. Figure~\ref{fig_er_largespec_and_hf} shows the amplitude of the transmitted microwave signal $|S_{21}(\omega)|$ as a function of the applied magnetic field from 0 to 77 mT. The input power of the microwave signal on the chip was set to be 1 fW, which corresponds to approximately 20 microwave photons inside each resonator. As anticipated from Fig.~\ref{fig_chip_and_crystal}~(a), the spectrum shows indeed both resonators coupled to the spin ensemble. The coupling manifests itself by a number of dispersive shifts and a large avoided level crossing (ALC) around 35~mT.

The regular pattern of 8 dispersive shifts is due to the HF transitions of the $^{167}$Er isotope with a nuclear spin of $I$~=~7/2. The coupling strength $v/2\pi$ and the inhomogeneous linewidth $\Gamma_2^{\star}/2\pi$ for each hyperfine transition coupled to LE resonator \#1 is summarized in table~\ref{Table1}.

\begin{table}[htb]
\begin{tabular}{|c|c|c|c|c|c|c|c|c|}
\hline
$m_I$ & 7/2 & 5/2 & 3/2 & 1/2 & $-$1/2 & $-$3/2 & $-$5/2 & $-$7/2 \\ \hline
$v/2\pi~$(MHz) & 9.7 & 7.3 & 7.6 & 7.2 & 3.8 & 5.7 & 5.7 & 5.0 \\ \hline
$\Gamma_2^{\star}/2\pi$~(MHz)& 31 & 27 & 29 & 26 & 15 & 22 & 33 & 25  \\ \hline
\end{tabular}
\caption{Nuclear spin projection $m_I$, coupling strengths $v/2\pi$ and linewidths $\Gamma_2^{\star}/2\pi$ for all 8 hyperfine transitions of the $^{167}$Er isotope coupled to LE resonator \#1.}\label{Table1}
\end{table}

Table~\ref{Table1} shows that the spin linewidth is 3 to 5 times larger than the coupling strength. It is expected that in the ideal crystal the inhomogeneous broadening is ultimately bounded by the presence of the nuclear $^{27}$Al spin bath with $I_{Al}=5/2$. In our setup, the magnetic field lies in the (001) plane of the crystal. Therefore, both magnetic classes are degenerate. However, any misalignment of the field out of this plane results in an additional inhomogeneous broadening~\cite{Bushev2011}. Interestingly, the coupling strength of the low field HF transitions are larger than for the high field ones, in contrast to $^{167}$Er:YSO~\cite{Probst2013}. The positions of the dispersive shifts in the spectrum do not coincide with the simulation presented on Fig.~\ref{fig_hf_simulation}. There is a discrepancy of at least 10\%. We refer that to the uncertainties in the determination of the hyperfine tensor \textrm{\textbf{A}}, which was also measured at the same level of accuracy at 12 Kelvin, see Refs.~\cite{Asatryan1997, Asatryan2002}.

\begin{figure}[htb]
\includegraphics[width=0.95\columnwidth]{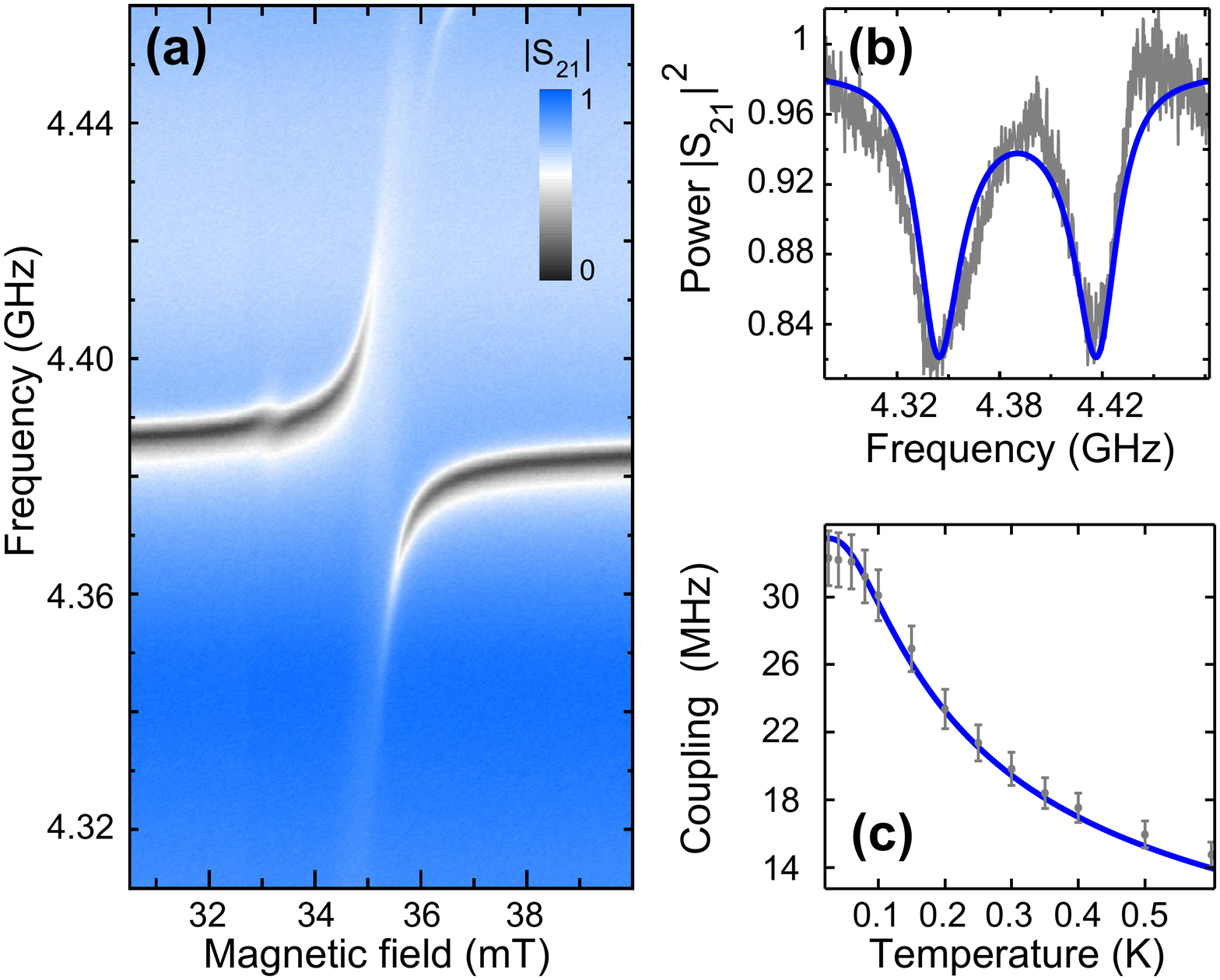}
\caption{(Color online)~\textbf{(a)}~Transmission spectrum of resonator \#1 strongly coupled to Er:YAP showing an avoided level crossing.~\textbf{(b)}~Extracted mode splitting (light gray) of the power spectrum $\left| S_{21}(\omega) \right|^{2}$ at 33~mT. The solid line shows the fit to the theory for zero detuning between the spins and cavity.~\textbf{(c)}~Temperature dependence of the coupling strength $v/2\pi$ from 25 to 600~mK. The light gray circles denote the data points and the solid line displays the fit to the theory.}
\label{fig_er_strong_coupling}
\end{figure}

The central avoided level crossing of resonator \#1 at a magnetic field of 33 mT is magnified in Fig.~\ref{fig_er_strong_coupling}~(a). In order to investigate the ALC, we remove the contribution of the transmission line, see~\cite{Probst2013} for the procedure. In Fig.~\ref{fig_er_strong_coupling}~(b) we plot the resulting power spectrum at 33 mT. A clear normal mode splitting confirms the coherent coupling of the LE resonator to the erbium electronic spin ensemble. The fit of the experimental data to the input-output theory model~\cite{Shuster2010} yields a coupling strength of $v/2\pi=32.2\pm0.4$~MHz and a linewidth of $\Gamma_2^{\star}/2\pi=16\pm1$~MHz. Compared to our experiments with the Er:YSO crystal~\cite{Probst2013}, the linewidth is only 25\% larger, while the coupling strength is approximately the same. We emphasize, that the resonant magnetic field of 33 mT is nearly one order of magnitude smaller than in the case of Er:YSO.

In order to investigate the properties of our electronic spin ensemble in greater detail, we perform a measurement of the coupling strength versus temperature, see Fig.~\ref{fig_er_strong_coupling}~(c). In contrast to the experiments with NV-centers in diamond~\cite{Majer2012}, our experimental data are in a good agreement with the model of an independent spin ensemble, given by $v=v_0[\tanh(\hbar \omega / 2k_B T)]^{1/2}$~\cite{Bushev2011}, where $v_0/2\pi$~=~33.5$~\pm~$0.3 MHz. For the temperature range from 25 to 150 mK a clear mode splitting is observed.

\begin{figure}[htb]
\includegraphics[width=0.95\columnwidth]{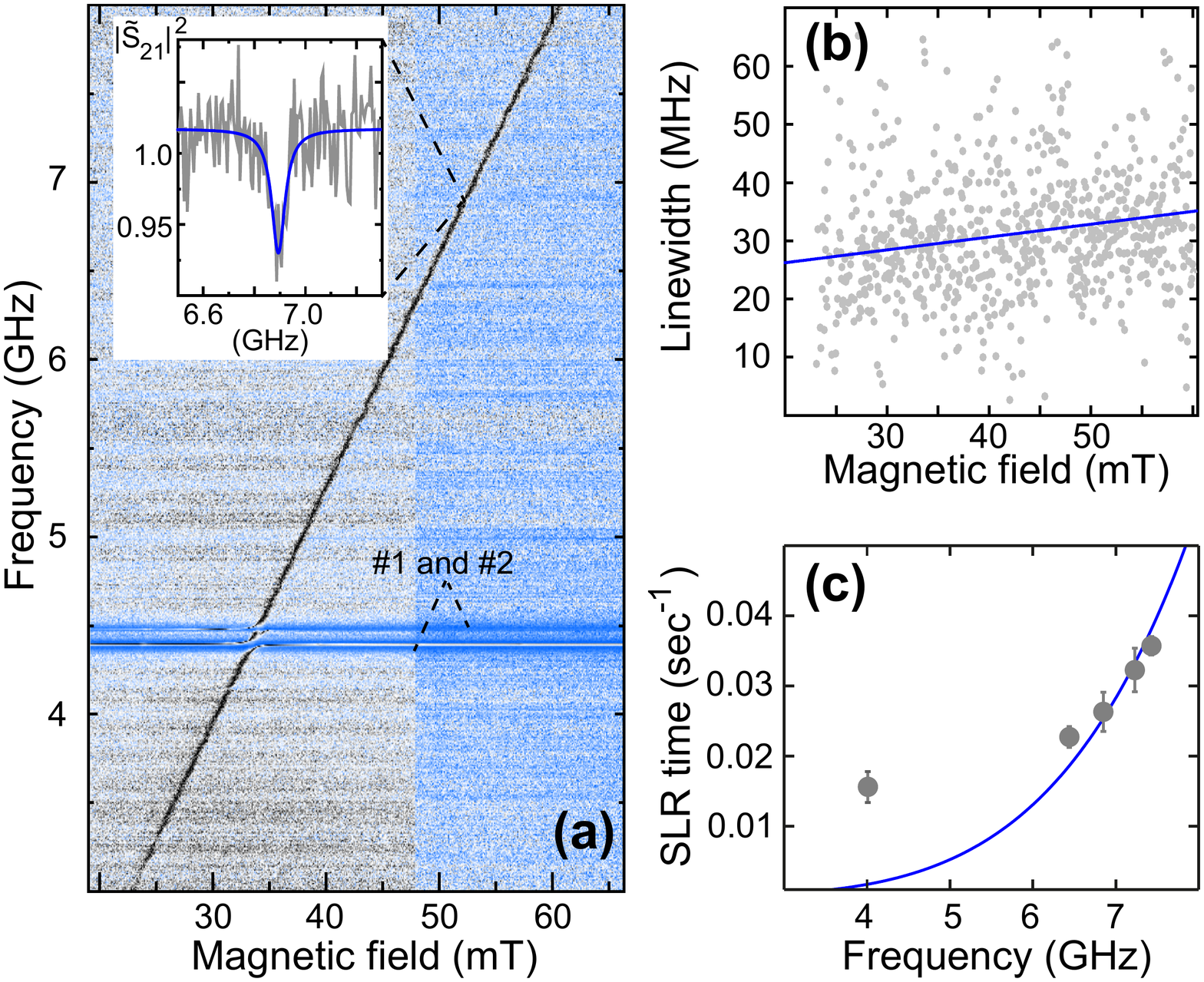}
\caption{(Color online)~\textbf{(a)}~Normalized ESR spectrum $|\tilde{S}_{21}(\omega)|^2$ of the Er:YAP crystal coupled to the 50~$\Omega$ transmission line. The absorption line is clearly visible on the spectrum and allows for the determination of the relaxation time $T_1$. The inset shows the typical microwave absorption profile of the Er:YAP crystal taken at 52 mT (gray line) and its fit to a Lorentzian.~\textbf{(b)}~Spin linewidth versus magnetic field.~\textbf{(c)}~Dependence of SLR time $T_1$ versus the spin resonance frequency (light gray dots). The solid line shows the fit to the theory.}
\label{fig_t1_absorptionline}
\end{figure}
In addition to the ALC, we observe a faint absorption line, which extends over the whole spectrum (Fig.~\ref{fig_t1_absorptionline}), and is caused by the direct absorption of microwaves travelling through the transmission line~\cite{Shuster2010, Probst2013}. Such a direct microwave absorption by a spin ensemble is of particular interest in the context of travelling wave quantum memories~\cite{Afzelius2013, Molmer2013, Probst2013}. Here, the presence of the absorption line allows us to study the relaxation mechanisms at millikelvin temperatures.

Figure~\ref{fig_t1_absorptionline}~(a) shows the normalized absorption power spectrum $|\tilde{S}_{21}(\omega)|^2=\left(|S_{21}(\omega,B)|/|S_{21}(\omega,\textrm{19~mT})|\right )^2$ in the range from 3 to 8 GHz corresponding to an applied magnetic field from 19 to 67 mT. The excitation power during the measurements was set to about 1 fW. The absorption line is clearly visible on the spectrum and corresponds to $\sim$~10$^{13}$ participating spins with a DC g-factor of 9.4~$\pm$~0.2.

The inset in the same figure demonstrates the typical absorption profile, which is measured at a field of 52 mT, and its fit to a Lorentzian. The magnitude of the absorption dip is approximately 8\% in the power spectrum and its HWHM linewidth is $\Gamma_2^{\star}/2\pi=31~\pm7$~MHz. From the shape of the absorption dip we extract the inhomogeneous linewidth for the field range from 23 to 60 mT, which is shown in Fig.~\ref{fig_t1_absorptionline}~(b). A linear fit of the data shows an increase of the linewidth with a rate of 0.23~MHz/mT. Such an increase can be explained by the inhomogeneity of the magnetic field in close proximity to the SC transmission line and mechanical stress. Small spatial variations of the magnitude of the \textrm{g}-factor in the crystal volume and a slightly lifted degeneracy associated with the two inequivalent ESR lines contribute to the broadening~\cite{Bushev2011}. Due to the large number of measurement points and the large measurement interval, the spin linewidth can be extrapolated to zero magnetic field with sufficient confidence. We obtain $\Gamma_2^{\star}/2\pi=22~\pm~1$~MHz for the intrinsic inhomogeneous broadening, which is close to the value obtained from coupling to the LE resonator, and it is also comparable to the inhomogeneous broadening measured for Er:YSO (12 MHz). The smaller values of the linewidth extracted from the coupling to the LE resonators could be explained in terms of the cavity protection effect~\cite{Diniz2011}.

The spin-lattice relaxation (SLR) time $T_1$ can be determined using the spin saturation technique described in Ref.~\cite{Probst2013}. Figure~\ref{fig_t1_absorptionline}~(c) presents the $T_1$ values plotted versus frequency. Close to the resonance frequencies of resonator \#1 and \#2, the SLR time cannot be extracted due to the interference with the resonator's signal. The SLR time ranges from half a minute to one minute at low frequencies. From the theory of paramagnetic ions in crystals~\cite{AbragamESR}, it is known that the $T_1$ time at low temperatures ($\hbar \omega \gg k_B T$) is dominated by a direct process, $(T_{1})^{-1}=R_d(\hbar\omega)^{5}\coth(\hbar\omega/2k_{B}T)$. However, this model cannot fully explain our experimental data, particularly at low frequencies, and requires an additional experimental study at millikelvin temperatures.

\begin{figure}[htb]
\includegraphics[width=0.95\columnwidth]{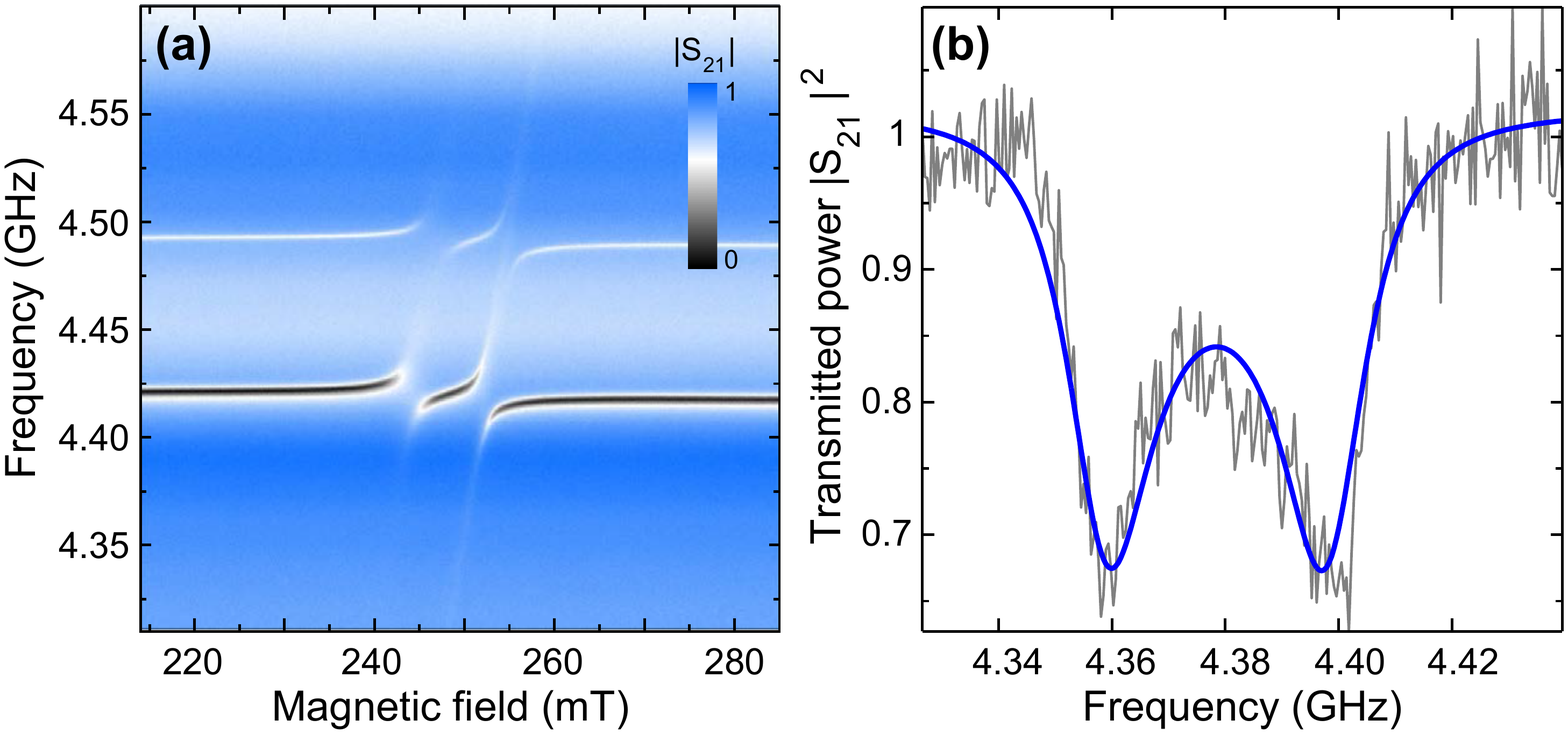}
\caption{(Color online)~\textbf{(a)}~Transmission spectrum of the Ce$^{3+}$:YAP spin ensemble magnetically coupled to LE resonators \#1 and \#2. Due to the misalignment of the field with respect to the crystal axis, two magnetically inequivalent transitions are observed.~\textbf{(b)}~Extracted mode splitting (light gray) of the power spectrum $\left| S_{21}(\omega) \right|^{2}$ at 252~mT, where the cerium spins and the LE cavity are in resonance. The solid line displays the fit to the theory.}
\label{fig_ce_strong_coupling}
\end{figure}

To our surprise, we found a significant presence of Ce$^{3+}$ impurities in the crystal, which reveal strong coupling at high magnetic fields. These ions were identified by their low g-factor of 1.25 and the absence of a HF structure. The ESR spectrum presented in Fig.~\ref{fig_ce_strong_coupling}~(a) shows one weakly and one strongly coupled electronic spin transition between 240 and 260 mT. The cerium ions distort the YAP crystal lattice to a greater extent because of their larger ionic radii as compared to erbium ions~\cite{Asatryan1997}. We believe that such a small distortion lifts the degeneracy of the two magnetically inequivalent sites for Ce$^{3+}$ in the (001) plane, thus two ESR transitions are detected. The deviation of the effective crystal axis for Ce:YAP with respect to Er:YAP is estimated to be about 0.2$^{\circ}$.

Figure~\ref{fig_ce_strong_coupling}~(b) displays the fit of the normal mode splitting to the second transition. The extracted coupling strengths for the first and the second transitions are $v/2\pi=15.2\pm0.1$~MHz and $v/2\pi=21.3\pm0.5$~MHz, respectively. The variation of the coupling strengths between both magnetic classes can be explained by a quite strong magnetic anisotropy of the Ce:YAP crystal with principal values of its \textrm{\textbf{g}}-tensor of $\textrm{g}_x=3.2$, $\textrm{g}_y=0.4$ and $\textrm{g}_z=0.4$, see Ref.~\cite{Asatryan1997}. Our measurement suggests the concentration of the Ce$^{3+}$ ions in the crystal to be about 0.005\%. The inhomogeneous spin linewidth extracted from the fit $\Gamma_2^{\star}/2\pi=15\pm1$~MHz is the same for both transitions.


In conclusion, we have presented an analysis of the first Er$^{3+}$:YAlO$_3$ based hybrid system for SC quantum circuits at millikelvin temperature. The spectrum consisting of electronic and hyperfine spin transitions fits within 10\% to the simulation. We demonstrate that the strong coupling regime for the electronic spin transition of the erbium ions can be attained at a relatively low magnetic field of approximately 33 mT. The intrinsic spin linewidth of Er:YAP was determined to be approximately 22 MHz. The spin relaxation time is on the order of one minute. In addition, we discovered the presence of Ce$^{3+}$ ions in the YAP crystal, which showed strong coupling, too. Our experiments demonstrate the promising potential of Er:YAP for its application in hybrid quantum systems, where low magnetic fields are desirable.

We thank A. Stockklauser and A. Wallraff for the stimulating discussions on the strong coupling regime, H. Majer-Flaig for the ESR simulation software and J.~H.~Cole and M.~Afzelius for the stimulating discussions. S.~P. acknowledges financial support by the LGF of Baden-W\"{u}rttemberg. This work was supported by the German Ministry of Education and Research (BMBF, project QUIMP).

\bibliographystyle{apsrev}

\bibliography{ErYALO_bib}

\end{document}